# Permanently densified SiO$_2$ glasses: a structural approach

C. Martinet, A. Kassir-Bodon, T. Deschamps, A. Cornet, S. Le Floch, V. Martinez and B. Champagnon

Université de Lyon, Université Lyon-1, UMR5306 CNRS, Institut Lumière Matière,

Bât. Kastler, 10 rue Ampère, 69662, Villeurbanne, France

E-mail : christine.martinet@univ-lyon1.fr

## Abstract

Densified silica can be obtained by different pressure and temperature paths and for different stress conditions, hydrostatic or including shear. The density is usually the macroscopic parameter used to characterize the different compressed silica samples. The aim of our present study is to compare structural modifications for silica glass, densified from several routes. For this, densified silica glasses are prepared from cold and high temperature (up to 1020°C) compressions. The different densified glasses obtained in our study are characterized by micro-Raman spectroscopy. Intertetrahedral angles from the main band relative to the bending mode decreases and their values are larger for densified samples from high temperature compression than those samples from cold compression. The relative amount of 3-membered rings deduced from the D$_2$ line area increases as a function of density for cold compression. The temperature increase during the compression process induces a decrease of the 3 fold ring population. Moreover, 3 fold rings are more deformed and stressed for densified samples at room temperature at the expense of those densified at high temperature. Temperature plays a main role in the reorganization structure during the densification and leads to obtaining a more relaxed structure with lower stresses than glasses densified from cold compression. The role of hydrostatic or non-hydrostatic applied stresses on the glass structure is discussed. From the Sen and Thorpe central force model, intertetrahedral angle average value and their distribution are estimated.

## 1.Introduction

SiO$_2$ glass under high pressure has been extensively studied both in experimental and theoretical studies and is a highly interesting material for many domains, such as solid state physics, the geosciences, and for technological applications. SiO$_2$ glass structure is a 3 dimensional network composed of connected SiO$_4$ units which form different sized rings, 33 to 10 fold rings, six fold rings being the most probable. 3 and 4 fold rings are much less probable, some percents compared to the total population of rings [1, 2]. However, their population evolution is very sensitive to thermal treatments or irreversible compression cycles. Furthemore, the proportion of 3 and 4 fold rings increases with the fictive temperature [3, 4, 5], after UV laser irradiation [6], neutron irradiation [7] or after an irreversible cold compression cycle [8,9].

Under high hydrostatic pressure at room temperature, $SiO_2$ glass has first a reversible behavior up to 9 GPa [9, 10, 11] and after that, an irreversible behavior above 9 GPa. The recovered glass, i.e. at ambient conditions, after an irreversible compression has got a different structure compared to the initial glass and the density progressively increases when the maximum applied pressure increases [12]. The limit of reversibility decreases when the glass is heated [13, 14, 15] or when shear stress is applied during compression [16]. For cold hydrostatic compression, above 20-25 GPa, the densification rate saturates at about 21% (density equal to 2.66 $g/cm^3$), determined from experimental [12, 17] and theoretical results [18]. During plastic deformation, irreversible changes occur and silica glass undergoes structural rearrangements at short-range and intermediate-range orders. Concerning short-range order, under high pressure, from X-ray diffraction measurements, silica structure evolves progressively from tetrahedral units to octahedral units, from 15 GPa to 42 GPa pressure range [19]. The Si coordination number increase could be certainly responsible of the densification process and the permanent structural modifications for $SiO_2$ glass. Nevertheless, the Si coordination number is fully reversible since, for the recovered glass, Si coordination returns to 4. Moreover, the absolute Raman scattering intensity decrease during compression is also observed from 10 GPa and is certainly related to the disappearance of the tetrahedral unit [20]. Short-range order is then unchanged for recovered silica glass. Nevertheless, in a recovered silica glass after an irreversible compression-decompression cycle, permanent modifications remain in the intermediate-range order. For plastic recovered silica, the intertetrahedral angles decrease and the ring statistics evolve compared to normal silica glass [9, 20, 21].

In-situ high pressure studies in the literature were performed mainly at room temperature with a Diamond Anvil Cell (D.A.C.) [20, 21]. Moreover, ex-situ studies after a compression process have been realized on silica in the past with simultaneous temperature and pressure being applied during the densification process using a Belt apparatus [16], a Bridgman apparatus [14], or using shock waves [22, 23]. Nevertheless, to our knowledge, no detail study has been performed to compare directly cold compression and compression at high temperature (up to 1020°C).

We then propose to detail here the temperature effects during a compression cycle and, in particular, a precise comparison between cold and high temperature compressions for different densities is performed. Several (P, T) paths are realized and the role of the temperature in the structural evolution in the intermediate-range order is discussed. For all these studies, ex-situ Raman spectroscopy is performed.

## 2.Experimental setup

In this study, silica samples were densified by several routes. For all experiments, pure $SiO_2$ glass (commercial "Suprasil 300") was used, with a low OH content, less than 1 ppm.

Room temperature densified samples were compressed from diamond anvil cell (D.A.C.), Chervin type, as it has been already detailed in a previous paper [24]. A mixture of methanol:ethanol:water (16:3:1) was used as the transmitting medium to insure hydrostatic compression conditions up to 15 GPa [25]. Ruby chips were introduced into the experimental volume in order to deduce the pressure, from the emission of $^2E$-$^4A_2$ ($R_1$ line) $Cr^{3+}$ transition [26]. Samples were compressed up to a maximal pressure, $P_{max}$, which was maintained for 60 min to insure that the equilibrium state was reached [27]. Several loadings on non-densified silica were performed with different $P_{max}$, above the elastic limit pressure, i.e. above 9 GPa, from 10 GPa to 26 GPa. The amples recovered from the DAC, after the compression-decompression cycle remain unbroken, due to negligible shear stresses during the loading cycle. The density was deduced from a calibration curve, relating the maximal attained pressures as a function of density obtained from recovered macroscopic silica samples [12]. This curve is available for densified silica from DAC experiments or multianvil press, i.e. in hydrostatic conditions, at room temperature.

Silica samples were also densified from high pressure and high temperature in a Belt Press [28]. Silica samples are cylinders (5.7 ± 0.1 mm in height and 3.95 ± 0.05 mm in diameter). Pyrophyllite used as transmitting medium doesn't insure fully hydrostatic conditions. Different pressure and temperature conditions were performed to obtain different densification rates. Belt press experiments consist in an increase of pressure up to the maximum desired value, and then to an increase in temperature up to the maximum desired value. Pressure and temperature are then kept constant for 10 minutes, after which the sample is cooled by stopping the heat supply. After this, the pressure decreases to atmospheric pressure in 20 minutes. Some non-hydrostatic stresses during the densification process induce a few sample fractures. The density for belt densified silica samples was measured using the flotation method based on Archimedes' principle.

Raman spectra were performed at room pressure in the backscattering configuration, with a Renishaw RM1000 micro-spectrometer, equipped with an edge filter and with a Peltier cooled CCD camera. The excitation is provided by a YAG: $Nd^{3+}$ laser line at 532 nm. All Raman measurements were performed after high pressure process, at atmospheric pressure and room temperature.

The different densified silica samples, their pressure and temperature process conditions and the measured densities are reported in table 1 both for DAC and Belt press experiments.

| Sample | Maximal pressure reached $P_{max}$ (GPa) | Temperature during compression (°C) | Density (g/cm$^3$) |
|---|---|---|---|
| Non densified | - | - | 2.20 |
| D.A.C.10 | 10 | 25 | 2.23 |
| D.A.C.12 | 12 | 25 | 2.30 |
| D.A.C.14 | 14 | 25 | 2.46 |
| D.A.C.16 | 16 | 25 | 2.51 |
| D.A.C.18 | 18 | 25 | 2.62 |
| D.A.C.26 | 26 | 25 | 2.66 |
| Belt_P4_T440 | 4 | 440 | 2.30 |
| Belt_P5_T440 | 5 | 440 | 2.40 |
| Belt_P4_T570 | 4 | 570 | 2.41 |
| Belt_P5_T750 | 5 | 750 | 2.54 |
| Belt_P5_T1020 | 5 | 1020 | 2.56 |

Table 1: (P,T) conditions and corresponding density for the densified samples

# 3. Results

Figure 1 shows Raman spectra of different densified silica glasses obtained from Belt press. From non-densified silica, the main band with a maximum located at around 438 cm$^{-1}$ corresponds to the bending mode of bridging oxygen $\nu_B$ (Si-O-Si) (O bending vibrations) and its large half width is attributed to the distribution of the Si-O-Si angles [29, 30]. The $D_1$, located at 490 cm$^{-1}$ and $D_2$, located at 606 cm$^{-1}$ "defect" lines are attributed to the breathing mode, corresponding to in-phase O-bending motion in four fold and three fold rings respectively [31].

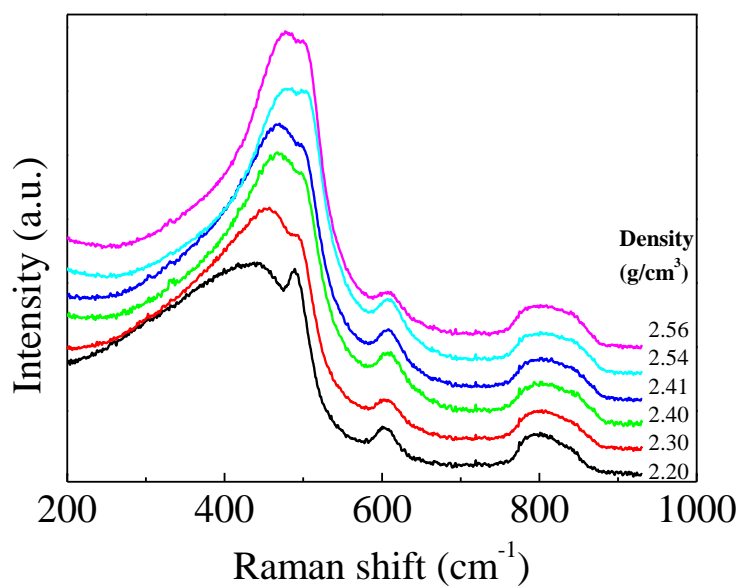

**Figure 1:** Raman spectra of densified silica glasses obtained from Belt press. Corresponding (P,T) conditions are detailed in table 1.

Figure 2 shows two Raman spectra for samples with similar densities from Belt Press and DAC both at 2.3 g/cm$^3$ (figure 2a) and 2.5 g/cm$^3$ (figure 2b) respectively. The Raman spectrum for the DAC sample with a density of 2.3 g/cm$^3$ shows larger $D_1$ and $D_2$ line intensities and higher main band frequency shift than that theRaman spectrum ofr the Belt sample with a similar density (figure 2a). For larger density (2.5 g/cm$^3$), both Raman spectra show large differences, in particular, the $D_2$ line area and main band position are dramatically different, with a different compression path (P,T). Moreover, for the DAC sample, the $D_1$ defect line and main band merges (figure 2b).

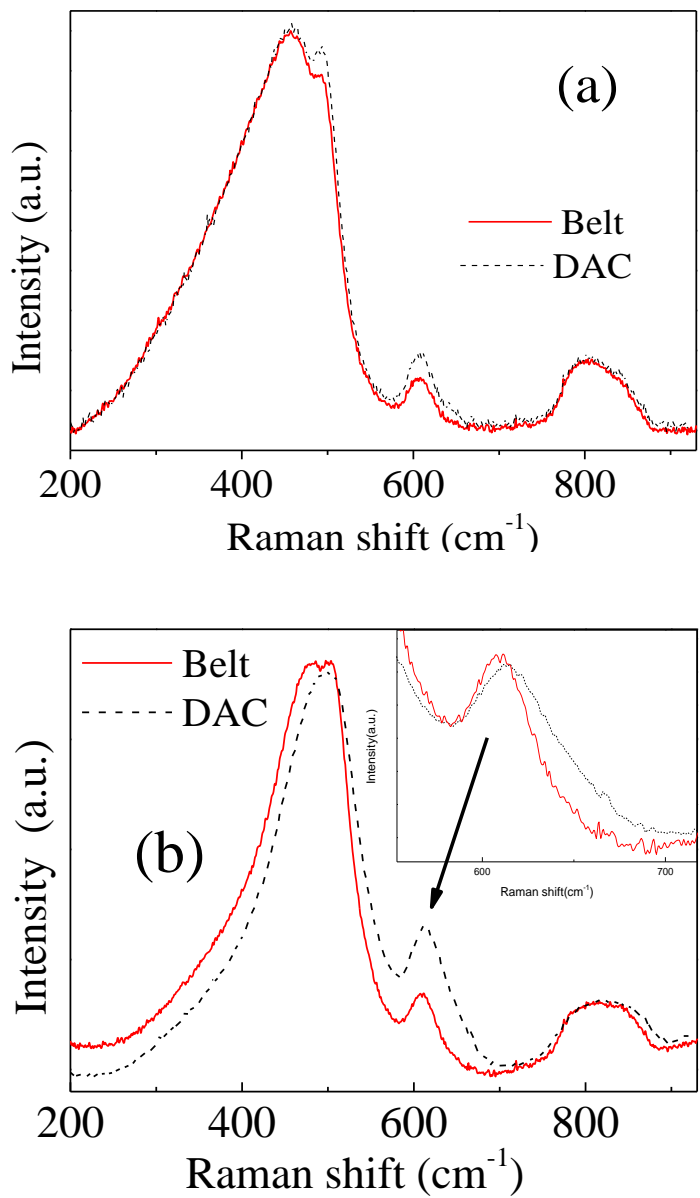

**Figure 2:** Raman spectra both of recovered densified glasses from Belt Press and from DAC, at similar densities (density equal to 2.3 g/cm$^3$, figure 2a and density equal to 2.5 g/cm$^3$, figure 2b). Red continuous line and black dashed line correspond to Raman spectra of Belt and DAC samples respectively.

Systematic determination of $D_2$ line area values are performed, normalized to the Raman spectrum area from 230 cm$^{-1}$ and 700 cm$^{-1}$. Figure 3a shows $D_2$ line area as a function of density both for recovered samples after cold and high temperature compressions. We observe that the $D_2$ line area increases when the density increases for cold compression, whereas, the D2 line area is density independent for high temperature compression.

Moreover, the $D_2$ line position shifts toward higher frequencies as a function of density whatever the pressure and temperature conditions are during the densification process (figure 3b). Nevertheless, $D_2$ line shifts faster as a function of density after a cold compression compared to high temperature compression (figure 3b). Main band maximum position and $D_1$ line shift toward higher frequencies as a function of the density for all samples but the variations are smaller for Belt press samples by comparison with DAC samples (figure 3c). The $D_1$ line and main band merge from densities of 2.45 – 2.50 g/cm$^3$, for cold compression. Then, in figure 3c), above these density values, only one position corresponding to the main band is determined. Moreover, the half width at half maximum of the main band decreases faster at the beginning of the densification and saturates above 2.5 g/cm$^3$, both after the cold and high temperatures compression-decompression cycles (figure 3d). This width is determined from the maximum of the main band shift and from the low frequency part, because the high frequency part of the band contains the $D_1$ line, which evolves in intensity and in position with the density.

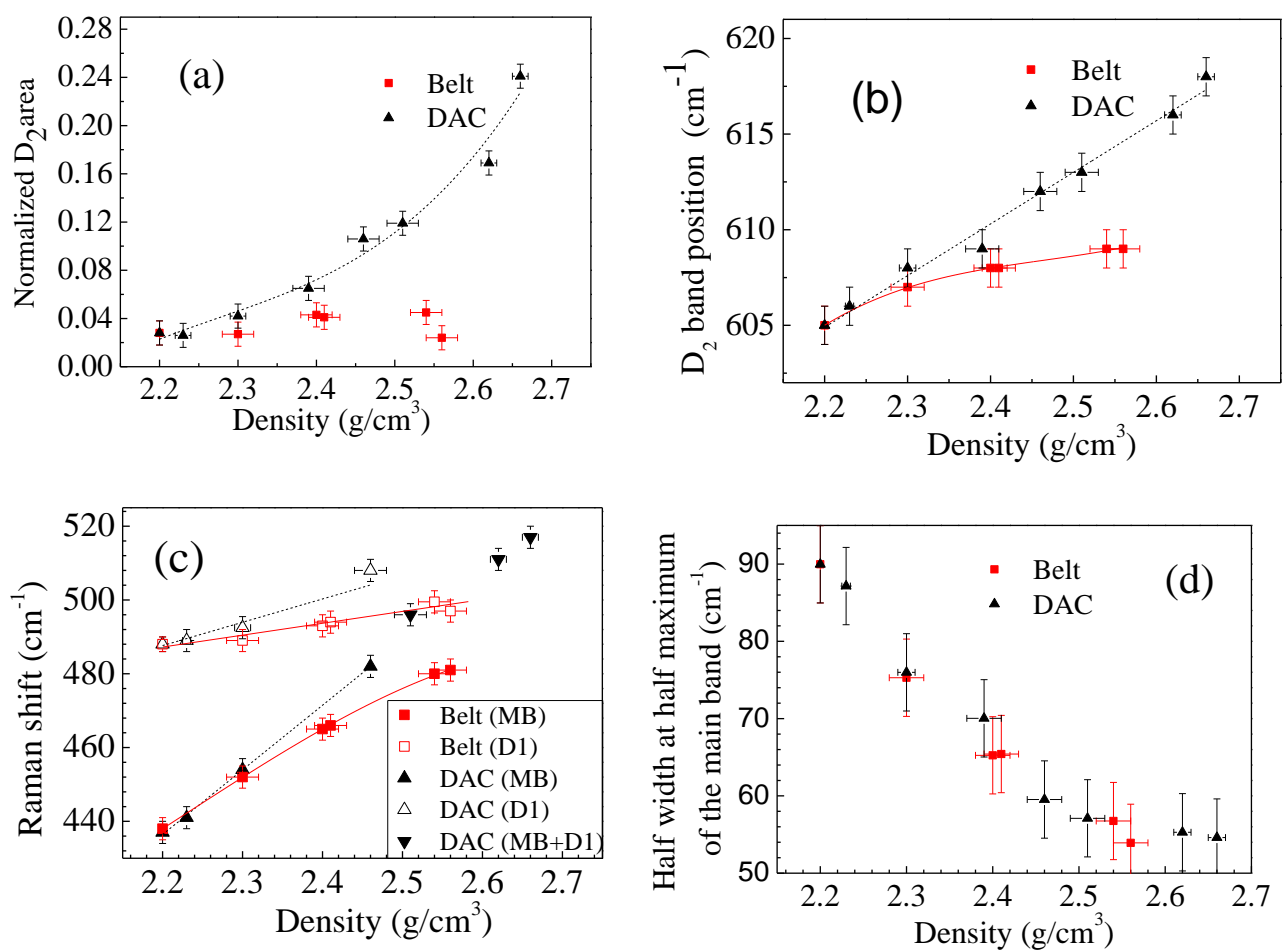

**Figure 3:** Normalized $D_2$ area (fig.3a), $D_2$ band shift (fig.3b), main band and $D_1$ positions (fig.3c) and half width at half maximum of the main band (fig. 3d) as a function of density, both for Belt samples (red squares) and for DAC samples (black triangles). Red continuous lines and black dashed lines are used for guidelines.

# 4. Discussion

In our present study, different (P,T) conditions allow us to obtain recovered silica samples with a progressive increasing density. To compare temperature effects on structural evolution, samples with different densities are prepared from cold and high temperature compressions.

Both in-situ and ex-situ compression experiments show a decrease both on Si-O-Si average and distribution angles for densified silica glass, deduced from Raman scattering [9, 20, 21] and from X-ray diffraction [32]. In our study, we observe that the main band maximum shifts to higher frequencies as a function of density, whatever pressure and temperature conditions. The main band can be directly connected to the Si-O-Si angular distribution using the Sen and Thorpe central force model [30]. The frequency is given by the following equation:

$$\omega = (2\alpha/m_o)^{(1/2)} (\cos\theta/2) \qquad (1)$$

where $\omega$ is the angular frequency corresponding to the Raman main band, $\alpha$ corresponds to the restoring central force constant between Si and O atoms, $m_o$ is the oxygen atom mass (in bending vibration, Si atoms are fixed and O atoms move), $\theta$ is the inter-tetrahedral angle. An intertetrahedral angle of 144° corresponds to the average standard position of the main band centred at 438 cm$^{-1}$ for non-densified silica. The restoring force $\alpha$ is assumed to be unchanged both for non-densified and densified silica glasses. Indeed, Si-O bond length is almost constant for recovered densified silica glasses [23, 29]. From equation (1), the main band shift to higher frequency corresponds to the Si-O-Si angles decreasing [9, 21].

From figure 3c, the main band shift as a function of density is larger for cold compression than for high temperature compression. Then, the intertetrahedral angle Si-O-Si decreases faster as a function of density for DAC samples in comparison to Belt samples. Therefore, for a similar density, intertetrahedral angles depend on the (P,T) densification path and differences increase at large densities. For example, from relation (1), intertetrahedral angle from the main band at the maximum intensity is 139.5 ± 0.3° and 140.7 ± 0.3° respectively for cold and high temperature densified glasses at 2.5 g/cm$^3$. For higher densities, it's difficult to evaluate the main band position due to its merging with the $D_1$ line. Differences in the intertetrahedral angle between high temperature and room temperature densification processes could be interpreted as a more strained structure after cold compression. A possible explanation is that compressed structure at room temperature is hindered to relax to a more stable structure. High temperatures during compression give a more relaxed structure and minimize stresses, in terms of intertetrahedral angles. Moreover, we observe an important intertetrahedral angle distribution decrease as a function of density whatever the compression path (P,T). In figure 3d, the half width at half maximum in the low frequency part is about 90 ±5 cm$^{-1}$ and 57 ± 5 cm$^{-1}$ for non-densified and 2.5g/cm$^3$ densified samples (DAC16 and Belt_P5_T750) respectively. From these values and relation (1), Si-O-Si angle distributions for non-densified and 2.5 g/cm$^3$ densified samples are of 7.6 ± 0.7 ° and 4.9 ± 0.7 ° respectively, for the low frequency part of the Raman spectra. The silica glass's structure after an irreversible compression is reorganized in a more homogeneous glass in terms of intertetrahedral angle distribution.

From the $D_1$ and $D_2$ line areas of Raman spectra, we can deduce the evolution of the relative population of four and three fold rings [31]. Molecular dynamics(MD) simulations have recently confirmed that the $D_1$ and $D_2$ lines are attributed to the breathing mode of 4 and 3 fold rings respectively and authors have established a direct proportionality between defect line areas and small tetrahedral cycles [33, 34]. Population rings evolution has been extensively studied after a densification process. Indeed, from the literature, the evolution of the ring size depends on the experimental conditions, DAC cold compression reveals an increase in the smaller ring population at the expense of larger rings, i.e. the 3 and 4 fold ring population increases after an irreversible compression, deduced from Raman scattering experiments [9, 20, 21]. For high temperature and high pressure, no evidence on the ring statistical evolution is observed after a densification process at 700°C and up to 8 GPa [13] whereas authors [14] have concluded to an increase of the smallest (3 and 4 fold) rings after a densification at 3.95 GPa and 530°C. Moreover, from simulation, no clear trend permits an outright conclusion. Indeed, after a densification process, some MD simulation results show that ring size statistics evolve to a proportional increase of larger rings at the expense of smaller rings [35] whereas from other MD simulation results, a proportional decrease of 5 and 6 fold ring, an increase of 3 and 4 fold rings and an increase population of larger rings (above 7 fold rings) have been deduced [36]. Thus, the role of temperature and hydrostatic pressure applied during compression is not clearly established. In our experiments, we observe that the $D_2$ line area intensity depends clearly on the densification path (figure3a). Indeed, for cold compression densified silica, the normalized $D_2$ line area increases continuously as a function of density (figure 3a) i.e. the relative proportion of three membered rings increases. For densities of 2.5-2.6 g/cm$^3$, the relative proportion of 3 fold rings is about three times larger for cold compression than for high temperature compression. Formation of 3 fold rings is then facilitated at room temperature and at high pressure.

For high temperature compressions, $D_2$ line area is density independent. 3 fold rings proportion is almost the same for the most densified sample at 1020°C and for non-densified silica. MD simulation at 1500K reveals the same trends with an increase of larger rings at the expense to smaller rings [35]. Our results and the theoretical results show that the temperature plays the main role in the structural evolution after an irreversible compression. Indeed, during high temperature compressions, thermal relaxations occur and the recovered structure corresponds to a more relaxed structure with an increase in the proportion of larger rings than for a recovered structure after cold compression. This result indicates that the 3 fold rings created during the densification process are unstable and temperature induces reorganization during the compression process with the formation of large rings at the expense of small ones.

From our experiments, the $D_2$ line position depends clearly on the densification path (P,T) (figure 3b). To quantify Si-O-Si angles in the 3 fold-rings, relation (1) can be used because the $D_2$ line corresponds to the breathing mode O-bending of small rings, these vibrations correspond mainly to a bending motion but also to a small stretching motion of the Si-O bond. Moreover, Si-O-Si angle values calculated from $D_2$ line are in good agreement with theoretical results [37]. This relation has been used in the past from Raman experimental results on densified silica glasses [29]. For example, from the maximum position of the $D_2$ line , Si-O-Si angles of 3 fold rings are 129.3±0.1° for non-densified silica and 129.0±0.1° and 128.6±0.1° for 2.5 g/cm$^3$ densified samples both from high temperature compression (Belt_P5_T750) and from cold compression (D.A.C.16). Si-O-Si angles decrease after a densification process and are lower for cold compression than for high temperature compression. This Si-O-Si angles decrease is correlated with the 3 fold ring puckering [38]. Moreover, the $D_2$ line width at half maximum clearly shows different values by comparing the non-densified silica, the densified silica samples at about 2.5 g/cm$^3$ both from high temperature compression (Belt_P5_T750) and from cold compression (D.A.C.16). Indeed, even if the $D_2$ line width value is unchanged both for non-densified silica and for Belt_P5_T750 sample (~ 27 cm$^{-1}$), on the contrary, the $D_2$ line width value is dramatically increased for the D.A.C.16 sample (~ 40 cm$^{-1}$). The $D_2$ line for the both densified samples can be visualized on the insert in the figure 2b and the Raman spectrum shows an obvious $D_2$ line broadening after a cold compression, leading in this last case to more strained 3 fold rings. To quantify distribution angles in the 3 fold rings, Si-O-Si angles are calculated from relation (1) and from the $D_2$ line width at half maximum. Distribution angles in the 3 fold rings are about 2.4 ± 0,3° for non-densified silica and 2.3 ± 0,3° and 3.6 ± 0,3° for Belt_P5_T750 and D.A.C.16 samples respectively. The maximum Si-O-Si angle and its width show clearly that 3-fold rings are more strained and stressed after a cold compression than high temperature compression and confirm the role of the temperature in the relaxation process during the densification even if the temperature is very low compared to glass transition temperatures of about 1250°C.

Even if temperature effects are put in evidence in our present results, we cannot exclude pressure applied effects on the recovered structure. In particular, to obtain the same density, the pressure applied should be greater for cold compression compared to high temperature compression [13, 14]. For example, the 2.5 g/cm$^3$ densified samples can be realized either from 5 GPa and 750°C (Belt_P5_T750) or from 16 GPa (D.A.C.16). In this last case, applied pressure is three times larger for cold compression than for high temperature compression. Moreover, recently, studies have been performed from MD simulation on silica under shear stress applied at low temperature and results have shown that the proportion of 3 fold rings increases clearly under shear stresses [39]. This conclusion is confirmed by our recent first results from Raman experiments on silica micro-indents where $D_2$ line area is more important in comparison with densified silica from hydrostatic cold compression for similar densities. These results confirmed the stress role (hydrostatic and shear) on the 3 fold ring creation at the expense of larger rings.

Non-densified silica and full densified recovered silica (i.e. 2.66 g/cm$^3$) are attributed to a low density amorphous (LDA) and a high density amorphous (HDA) phases respectively [24, 40]. This transition needs thermal activation energy due to energetic barrier between these two energy minima. Nevertheless, pressure induces mechanical instabilities leading to a barrier decrease between the two energy minima [41]. Each megabasin (LDA/HDA) is composed of many local minima [42]. These different structural states, observed in our study, depending on the densification path, correspond to probably different local energy minima. Indeed, during irreversible (P,T) cycle, temperature plays a main role in the formation of a more stable structure which is energetically lower than for a cold compression [43]. The temperature permits the system to relax to a secondary potential energy minimum and the structure is then probably more stable. This hypothesis should be confirmed by thermal relaxation process on densified glass.

## 5. Conclusion

Studies both on the influence of temperature and pressure on densified silica glasses are detailed here from ex-situ Raman experiments. In particular, 3 fold rings increase as a function of density for cold compression whereas after a high temperature compression, 3 fold ring is almost density independent. Thus, silica glass densified at high pressure and high temperature have larger intertetrahedral angles and lower proportion of 3 fold rings compared to cold compression. The $D_2$ line width is also significantly larger for cold compression showing that the 3 fold rings are more deformed in this case. Temperature permits to the system to relax to a secondary potential energy minimum and the structure is then probably more stable after a high temperature compression.

## Acknowledgments

The authors would like to thank to A. Tanguy (Institut Lumière Matière –Université Lyon1-France) for fruitful discussions. The authors would like to thank to N. Blanchard, native English speaker, to corrected proofs. Raman experiments were performed at the vibrational spectroscopies Platform at University Lyon 1-France (CECOMO) and the Belt press experiments at the Lyon Platform of Experiments under Extreme Conditions (PLECE). Present work was supported by A.N.R. MECASIL, French program facility.

## References

[1] Rino J P, Ebbsjö I 1993 *Phys. Rev. B* **47** 3053

[2] Jin W, Kalia R K, Vashishta P and Rino J P 1994 *Phys. Rev. B* **50** 118

[3] Geissberger A E and Galeener F L 1983 *Phys. Rev. B* **28** 3266

[4] Tomozawa M, Hong J W and Ryu S R 2005 *J. Non-Cryst. Solids* **351** 1054

[5] Martinet C, Martinez V, Coussa C, Champagnon B and Tomozawa M 2008 *J. Appl. Phys.* **103** 083506

[6] Borelli N F, Smith C, Allan D C and Seward T P 1997 *J. Opt. Soc. Am.* **14** 1607

[7] Devine R A B 1994 *Nucl. Instr. Meth. Phys. Res. B* **91** 378

[8] Sugiura H and Yamadaya T 1992 *J. Non-Cryst. Solids* **144,** 151

[9] Champagnon B, Martinet C, Boudeulle M, Vouagner D, Coussa C, Deschamps T, Grosvalet L 2008 *J. Non-Cryst. Solids* **354** 569

[10] Bridgman P W and Simon I 1953 *J. Appl. Phys.* **24** 405

[11] Walrafen G E and Krishnan P N 1981 *J. Chem. Phys.* **74** 5328

[12] Rouxel T, Ji H, Hammouda T and Moréac A 2008 *Phys. Rev. Lett.* **100** 225501

[13] Poe B, Romano C and Henderson G 2004 *J. Non-Cryst. Solids* **341** 162

[14] McMillan P, Piriou B and Couty R 1984 *J. Chem. Phys.* **81** 4234

[15] Devine R A B, Dupree R, Farnan I and Capponi J J 1987 *Phys. Rev. B* **35** 2560

[16] Mackenzie J D 1963 *J. of Am. Ceram. Soc.* **46** 461

[17] Polian A and Grimsditch M 1993 *Phys. Rev. B* **47** 13979

[18] Della Valle R G and Venuti E 1996 *Phys. Rev. B* **54** 3809

[19] Benmore C J, Soignard E, Amin S A, Guthrie M, Shastri S D, Lee P L and Yarger J L 2010 *Phys. Rev. B* **81** 054105

[20] Polsky C H, Smith K H and Wolf G H 1999 *J. Non-Cryst. Solids* **248** 159

[21] Hemley R J, Mao H K, Bell P M and Mysen B O 1986 *Phys. Rev. Lett.* **57** 747

[22] Sugiura H, Ikeda R, Kondo K and Yamadaya T 1997 *J. of Appl. Phys.* **81** 1651

[23] Shimada Y, Okuno M, Syono Y, Kikuchi M, Fukuoka K and Ishizawa N 2002 *Phys. and Chem. of Min.* **29** 239

[24] Deschamps T, Kassir-Bodon A, Sonneville C, Margueritat J, Martinet C, De Ligny D, Mermet A and Champagnon B 2013 *J. Phys. Cond. Mat.* **25** 025402

[25] Klotz S, Chervin J C, Munsch P and Le Marchand G *J.* 2009 *Phys. D: Appl. Phys.* **42** 075413

[26] Piermarini G J, Block S, Barnett J D and Forman R A 1975 *J. Appl. Phys.* **46** 2774

[27] Sonneville C, Mermet A, Champagnon B, Martinet C, Margueritat J, De Ligny D, Deschamps T and Balima F *J.* 2012 *Chem. Phys.* **137** 124505

[28] Hall H T 1960 *Rev. Sci. Instrum.* **31** 125

[29] Hehlen B 2012 *J. Phys. Cond. Mat.* **22** 025401

[30] Sen P N and Thorpe M F 1977 *Phys. Rev. B* **15** 4030

[31] Galeener F L 1982 *J. Non-Cryst. Solids* **49** 53

[32] Meade C, Hemley R J and Mao H K 1992 *Phys. Rev. lett.* **69** 1387

[33] Pasquarello A and Car R 1998 *Phys. Rev. Lett.* **80** 5145

[34] Burgin J, Guillon C, Langot P, Vallée F, Hehlen B and Foret M 2008 *Phys. Rev. B* **78** 184203

[35] Huang L and Kieffer J 2004 *Phys. Rev. B* **69** 224204

[36] Davila L, Caturla M, Kubota A, Sadigh B, Dıaz de la Rubia T, Shackelford J F, Risbud S H and Garofalini S H 2003 *Phys. Rev. Lett.* **91** 205501

[37] Umari P, Gonze X and Pasquarello A 2003 *Phys. Rev. lett.* **90** 027401

[38] Barrio R A, Galeener F L, Martinez E and Elliott R 1993 *Phys. Rev. B* **48** 15672

[39] Shcheblanov N S, Mantisi B, Umari P and Tanguy A *Phys. Rev. B (submitted)*

[40] Sonneville C, Deschamps T, Martinet C, De Ligny D, Mermet A and Champagnon B 2013 *J. Non-Cryst. Solids* **382** 133

[41] Lacks D J 1998 *Phys. Rev. Lett.* **80** 5385

[42] Lacks D J 2000 *Phys. Rev. Lett.* **84** 4629

[43] Liang Y, Miranda C R and Scandolo S 2007 *Phys. Rev. B* **75** 024205